\documentstyle[twocolumn,seceq,epsbox]{jpsj}

\def\runtitle{Field-Induced Transverse N\'{e}el Ordering in TlCuCl$_3$}
\def\runauthor{Hidekazu {\sc Tanaka} {\it et al.}}

\title
{
Observation of Field-Induced Transverse N\'{e}el Ordering\\ 
in the Spin Gap System TlCuCl$_3$
}

\author
{ 
Hidekazu {\sc Tanaka}\footnote{E-mail: tanaka@lee.phys.titech.ac.jp.}, Akira {\sc Oosawa}, Tetsuya {\sc Kato}$^{1}$ Hidehiro {\sc Uekusa}$^{2}$, Yuji {\sc Ohashi}$^{2}$, Kazuhisa {\sc Kakurai}$^{3}$ and Andreas {\sc Hoser}$^{4}$
}

\inst
{
Department of Physics, Tokyo Institute of Technology, Oh-okayama, Meguro-ku, Tokyo 152-8551\\
$^{1}$Faculty of Education, Chiba University, Inage-ku, Chiba 263-8522\\
$^{2}$Department of Chemistry, Tokyo Institute of Technology, Oh-okayama, Meguro-ku, Tokyo 152-8551\\
$^{3}$Neutron Scattering Laboratory, Institute for Solid State Physics,
The University of Tokyo, Tokai, Ibaraki 319-1106\\
$^{4}$Hahn-Meitner-Insitut, Glienicker Stra{\ss}e 100, D--14109 Berlin, Germany
}

\recdate
{
\hspace{7cm}}

\abst
{
Neutron elastic scattering experiments have been performed on the spin gap system TlCuCl$_3$ in magnetic fields parallel to the $b$-axis. The magnetic Bragg peaks which indicate the field-induced N\'{e}el ordering were observed for magnetic field higher than the gap field $H_{\rm g}\approx 5.5$ T at $Q=(h, 0, l)$ with odd $l$ in the $a^*-c^*$ plane. The spin structure in the ordered phase was determined. The temperature and field dependence of the Bragg peak intensities and the phase boundary obtained were discussed in connection with a recent theory which describes the field-induced N\'{e}el ordering as a Bose-Einstein condensation of magnons.
}

\kword
{
TlCuCl$_3$, spin gap, field-induced magnetic ordering, spin structure, neutron elastic scattering, Bose-Einstein condensation of magnons
}

\begin{document}
\sloppy
\maketitle

The singlet ground state with the excitation gap (spin gap) is a notable realization of the macroscopic quantum effect in quantum spin systems. When a magnetic field is applied in the spin gap system, the gap $\Delta$ is suppressed and closes completely at the gap field $H_{\rm g}=\Delta/g\mu_{\rm B}$. For $H>H_{\rm g}$ the system can undergo magnetic ordering due to three-dimensional (3D) interactions. Such field-induced magnetic ordering was studied first for Cu(NO$_3$)$_2 \cdot \frac{5}{2}$H$_2$O \cite{Diederix}. However, the magnetic properties near $H=H_{\rm g}$ have not been investigated because of the very low ordering temperature, the maximum of which is 0.18 K. Recently, the study of field-induced magnetic ordering has been revived, because the spin gap has been found in many quantum spin systems. Field-induced 3D ordering has been observed in several quasi-one-dimensional spin gap systems \cite{Honda,Hammer,Chaboussant,Hagiwara,Manaka}. \par
	This paper is concerned with field-induced magnetic ordering in TlCuCl$_{3}$. This compound has a monoclinic structure (space group $P2_1/c$) \cite{Takatsu}. TlCuCl$_3$ contains planar dimers of Cu$_2$Cl$_6$, in which Cu$^{2+}$ ions have spin-$\frac{1}{2}$. These dimers are stacked on top of one another to form infinite double chains parallel to the crystallographic $a$-axis. These double chains are located at the corners and center of the unit cell in the $b-c$ plane, and are separated by Tl$^+$ ions. The magnetic ground state is the spin singlet with the excitation gap $\Delta /k_{\rm B}\approx 7.5$ K \cite{Shiramura,Oosawa1}. The magnetic excitations in TlCuCl$_{3}$ were investigated by Oosawa {\it et al.} \cite{Oosawa2}, who found that the lowest excitation occurs at $Q=(0, 0, 1)$ and its equivalent reciprocal points, as observed in KCuCl$_3$ \cite{Kato,Cavadini}. The origin of the gap is the strong antiferromagnetic interaction $J=5.26$ meV on the planar dimer Cu$_2$Cl$_6$ in the double chain. The neighboring dimers couple magnetically along the chain and in the $(1, 0, -2)$ plane. \par
	Our previous magnetic measurements revealed that TlCuCl$_{3}$ undergoes 3D magnetic ordering in magnetic fields higher than the gap field $H_{\rm g}\approx 5.5$ T \cite{Oosawa1}. The magnetization exhibits a cusplike minimum at the ordering temperature $T_{\rm N}$. The phase boundary on the temperature vs field diagram is independent of the field direction when normalized by the $g$-factor, and can be represented by the power law
$$\left( g/2 \right) \left[H_{\rm N}(T)-H_{\rm g}\right] \propto T^{\phi}\ ,\eqno(1)$$ 
with $\phi=2.2$, where $H_{\rm N}(T)$ is the transition field at temperature $T$. These features cannot be explained by the mean-field theory \cite{Tachiki1,Tachiki2}. \par
	Nikuni {\it et al.} \cite{Nikuni} argued that field-induced magnetic ordering in TlCuCl$_{3}$ can be represented as a Bose-Einstein condensation (BEC) of excited triplets (magnons). In the magnon BEC theory, the external field $H$ and the total magnetization $M$ are related to the chemical potential $\mu$ and the number of magnons $N$, respectively, as $\mu =g{\mu}_{\rm B}(H-H_{\rm g})$ and $M=g{\mu}_{\rm B}N$. The observed temperature dependence of the magnetization and the field dependence of the ordering temperature \cite{Oosawa1} were qualitatively well described by the magnon BEC theory based on the Hartree-Fock (HF) approximation \cite{Nikuni}. \par
	If the magnons undergo BEC at ordering vector $Q_0$ for $H>H_{\rm g}$, then the transverse spin components have long-range order, which is characterized by the same wave vector $Q_0$. When the magnetization is small, {\it i.e.,} magnons are dilute, the transverse magnetization $m_{\perp}$ per site is expressed by
$$m_{\perp}=g{\mu}_{\rm B}\sqrt{n_{\rm c}/2}\ ,\eqno(2)$$
where $n_{\rm c}$ is the condensate density. Thus the condensate density $n_{\rm c}$ is proportional to the magnetic Bragg peak intensity. \par
\begin{table}[tbp]
\caption{Atomic coordinates of TlCuCl$_3$.}
\label{table:1}
\begin{center}
\begin{tabular}{@{\hspace{\tabcolsep}\extracolsep{\fill}}cccc}
\hline
  & $x$ & $y$ & $z$ \cr
\hline
Tl	& 0.7780 & 0.1700 & 0.5529 \cr
Cu	& 0.2338	& 0.0486 & 0.1554 \cr
Cl(1) & 0.2656	& 0.1941 & 0.2597\cr 
Cl(2) & 0.6745	& $-$0.0063 & 0.3177\cr 
Cl(3) & $-$0.1817	& 0.0966 & $-$0.0353\cr 
\hline
\end{tabular}
\end {center}
\end{table}
	In order to confirm the transverse spin ordering corresponding to the off-diagonal long-range order of the BEC, and to investigate the temperature and field dependence of the transverse magnetization $m_{\perp}$ corresponding to the condensate density $n_{\rm c}$, we performed neutron elastic scattering experiments on TlCuCl$_3$. \par
	The details of sample preparation were reported in reference \citen{Oosawa1}. Since the precise structural analysis for TlCuCl$_3$ has not been reported, we performed the single-crystal X-ray diffraction. The lattice parameters at room temperature were determined as $a=3.9815 \rm{\AA}$, $b=14.1440 \rm{\AA}$, $c=8.8904 \rm{\AA}$ and $\beta=96.32^{\circ}$. The atomic coordinates obtained are shown in Table I. \par
	Neutron scattering experiments were carried out at the E1 spectrometer of the BER II Research Reactor of the Hahn-Meitner Institute with the vertical field cryomagnet VM1. The incident neutron energy was fixed at $E_{\rm i}=13.9$ meV, and the horizontal collimation sequence was chosen as 40'-80'-40'. A single crystal of $\sim 0.4$ cm$^3$ was used. The sample was mounted in the cryostat with its (0, 1, 0) cleavage plane parallel to the scattering plane, so that the reflections in the $a^*-c^*$ plane were investigated. The sample was cooled to 0.05 K using a dilution refrigerator. An external magnetic field of up to 12 T was applied along the $b$-axis. Lattice parameters $a=3.899$ $\rm{\AA}$, $c=8.750$ $\rm{\AA}$ and $\beta=95.55^{\circ}$ were used at helium temperatures. \par
	Magnetic Bragg reflections were observed for $H>H_{\rm g}\approx 5.5$ T, which could be indexed by $Q=(h, 0, l)$ with odd $l$. Figure 1 shows the ${\theta}-2{\theta}$ scans for $(1, 0, -3)$ reflection measured at $H=0$ and 12 T at 1.9 K. These reciprocal lattice points are equivalent to those for the lowest magnetic excitation at zero field \cite{Oosawa2}. This result indicates that the magnetic unit cell is the same as the chemical one. Ferromagnetic Bragg reflections due to the induced moment could not be detected in the present measurements. The integrated intensities of nine magnetic reflections were measured at 1.9 K under an external field of 12 T. The results are summarized in Table II together with the calculated intensities. The intensities of nuclear Bragg reflections were also measured to determine the magnitude of the magnetic moment per site. The nuclear peaks were observed at $Q=(h, 0, l)$ with even $l$, as expected from the space group $P2_1/c$. However, very weak nuclear peaks were also observed for odd $l$. To refine the magnetic structure, we used the atomic coordinates shown in Table I and the nuclear scattering lengths $b_{\rm Tl}=0.878$, $b_{\rm Cu}=0.772$ and $b_{\rm Cl}=0.958$ with the unit of 10$^{-12}$ cm \cite{Sears}. The magnetic form factors of Cu$^{2+}$ were taken from reference \citen{Brown}. The extinction effect was evaluated by comparing observed and calculated intensities for various nuclear Bragg reflections. \par
\begin{figure}[tbp]
  \begin{center}
    \epsfigure{file=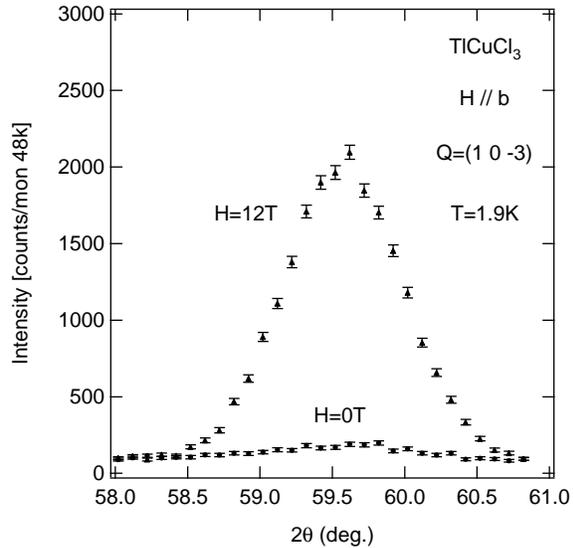,width=80mm}
  \end{center}
  \caption{${\theta}-2{\theta}$ scans for $(1, 0, -3)$ reflection measured at $H=0$ and 12 T at 1.9 K.}
  \label{fig:1}
\end{figure}
\begin{table}[t]
\caption{Observed and calculated magnetic Bragg peak intensities at $T=1.9$ K and $H=12$ T for $H{\parallel}b$. The intensities are normalized to the (0, 0, 1)$_{\rm M}$ reflection. $R$ is the reliability factor given by $R=\sum_{h,k,l}|I_{\rm cal}-I_{\rm obs}|/\sum_{h,k,l}I_{\rm obs}$.}
\label{table:2}
\begin{center}
\begin{tabular}{@{\hspace{\tabcolsep}\extracolsep{\fill}}ccc}
\hline
$(h, k, l)$ & $I_{\rm obs}$ & $I_{\rm cal}$ \cr
\hline
(0, 0, 1)$_{\rm M}$ & 1 & 1 \\
(0, 0, 3)$_{\rm M}$ & 0 & 0.014 \\
(0, 0, 5)$_{\rm M}$ & 0.063 & 0.107 \\
(1, 0, 1)$_{\rm M}$ & 0.031 & 0.009 \\
(1, 0, $-1$)$_{\rm M}$ & 0.074 & 0.127 \\
(1, 0, 3)$_{\rm M}$ & 0.056 & 0.024 \\
(1, 0, $-3$)$_{\rm M}$ & 0.398 & 0.380 \\
(2, 0, 1)$_{\rm M}$ & 0.012 & 0.013 \\
(2, 0, $-1$)$_{\rm M}$ & 0.128 & 0.102 \\
\hline
$R$ & 0.12 \\
\hline
\end{tabular}
\end {center}
\end{table}
	Figure 2 shows the spin structure determined at 1.9 K and 12 T. The directions of spins are in the $a-c$ plane which is perpendicular to the applied field. Spins on the same dimers represented by thick lines in Fig. 2 are antiparallel. Spins are arranged in parallel along a leg in the double chain, and make an angle of 39$^{\circ}$ with the $a$-axis. The spins on the same legs in the double chains located at the corner and the center of the unit cell in the $b-c$ plane are antiparallel. Although scans in the reciprocal plane including the $b^*$-axis were not performed, the spin structure shown in Fig. 2 is uniquely determined within the present measurements. Thus the transverse magnetic ordering predicted by the theory \cite{Tachiki1,Tachiki2,Nikuni} was confirmed. \par
	Comparing magnetic peak intensities with those of nuclear reflections, we evaluate the magnitude of the transverse magnetization at 1.9 K and 12 T as $m_{\perp}=g{\mu}_{\rm B}{\langle}S_{\perp}{\rangle}=0.26(2)$ ${\mu}_{\rm B}$. Since the N\'{e}el temperature for $H=12$ T is $T_{\rm N}=7.2$ K, the magnitude of $m_{\perp}$ at 1.9 K must be very close to that at zero temperature (see Fig. 2). For $T\rightarrow 0$, the condensate density $n_{\rm c}$ approximates the density of the magnon $n$, which is given by $n=2m/g{\mu}_{\rm B}$ with the magnetization $m$ per site. The value of $n$ at 12 T and 1.8 K is obtained as $n\approx 0.028$ with $m\approx 0.03\ {\mu}_{\rm B}$ and $g=2.06$ \cite{Shiramura,Oosawa1}. Substituting $n_{\rm c}\approx n\approx 0.028$ into eq. (2), $m_{\perp}\approx 0.24$ ${\mu}_{\rm B}$. This value agrees with $m_{\perp}=0.26(2)$ ${\mu}_{\rm B}$ obtained by the present measurement.  \par
\begin{figure}[tbp]
  \begin{center}
    \epsfigure{file=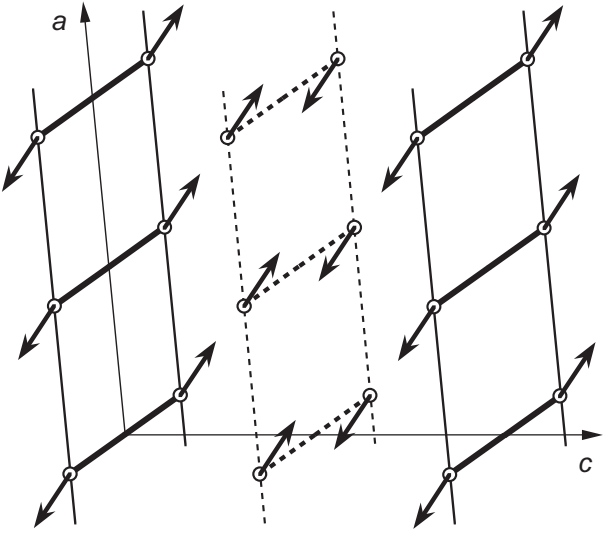,width=80mm}
  \end{center}
  \caption{Spin structure in the ordered phase of TlCuCl$_3$. The external field is applied along the $b$-axis. The double chain located at the corner and the center of the chemical unit cell in the $b-c$ plane are represented by solid and dashed lines, respectively.}
  \label{fig:2}
\end{figure}
	Figure 3 shows the temperature dependence of the magnetic peak intensity for $(1, 0, -3)$ measured in various magnetic fields. On the right abscissa, we show $m_{\perp}^2$ which corresponds to the condensate density $n_{\rm c}=2\left(m_{\perp}/g{\mu}_{\rm B}\right)^2$. The N\'{e}el ordering can be detected at the temperatures indicated by the thick arrows. With increasing external field, the ordering temperature $T_{\rm N}(H)$ and the intensity of the magnetic Bragg peak increase. The anomaly around $T_{\rm N}(H)$ becomes smeared as the magnetic field approaches $H_{\rm g}$, which leads to the difficulty in determining the N\'{e}el temperature from the temperature dependence of the Bragg peak intensity for $H\leq 6.6$ T. Since the sharp anomaly is observed at the transition field $H_{\rm N}(T)$ in the field dependence of the Bragg intensity (see Fig. 4), and the sharp phase transition was detected through the specific heat measurement \cite{Oosawa3}, the smearing of the peak intensity around $T_{\rm N}(H)$ is attributed not to the intrinsic smearing of the ordering temperature due to additional effects such as the staggered inclination of the principal axis of the {\it g}-tensor and the Dzyaloshinsky-Moriya interaction, but to the diffuse scattering. The amount of diffuse scattering around $T_{\rm N}(H)$ seems to be independent of peak intensity at $T=0$. \par
\begin{figure}[tbp]
  \begin{center}
    \epsfigure{file=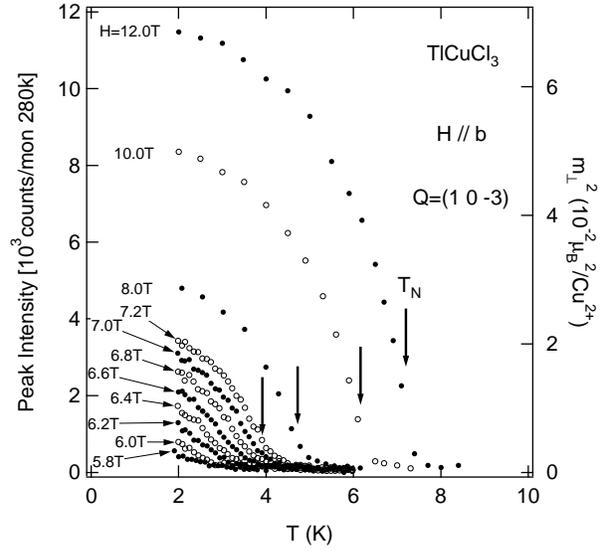,width=80mm}
  \end{center}
  \caption{Temperature dependence of the Bragg peak intensity for $(1, 0, -3)$ reflection in various magnetic fields. The square of the transverse magnetization per site $m_{\perp}^2$ is shown on the right abscissa.}
  \label{fig:3}
\end{figure}
\begin{figure}[tbp]
  \begin{center}
    \epsfigure{file=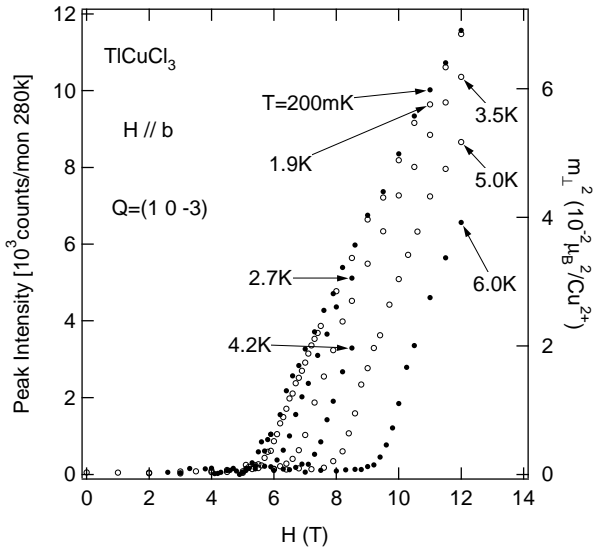,width=80mm}
  \end{center}
  \caption{Magnetic field dependence of the Bragg peak intensity for $(1, 0, -3)$ reflection at various temperatures. The square of the transverse magnetization per site $m_{\perp}^2$ is shown on the right abscissa.}
  \label{fig:4}
\end{figure} 
	Figure 4 shows the field dependence of the Bragg intensity for $(1, 0, -3)$ reflection measured at various temperatures. On the right abscissa, we show $m_{\perp}^2$. The intensity remains almost zero up to the transition field $H_{\rm N}(T)$, and then increases rapidly. It is clear that field-induced N\'{e}el ordering takes place at $H_{\rm N}(T)$. At 0.2 K, the intensity ($m_{\perp}^2$) is almost proportional to $H-H_{\rm g}$ just above $H_{\rm g}$, and is slightly concave at higher fields. With increasing temperature, the transition field $H_{\rm N}(T)$ increases, and the slope just above $H_{\rm N}(T)$ increases. \par
	The magnon BEC theory based on the HF approximation \cite{Nikuni} predicts that the magnon density $n$ and the condensate density $n_{\rm c}$ are both concave functions of $\mu =g{\mu}_{\rm B}(H-H_{\rm g})$ for $T{\rightarrow}0$ \cite{Nikuni}. The deviation from the linear function of $\mu$ is due to the  noncondensate magnons, and is proportional to ${\mu}^{3/2}$. The field dependence of the Bragg intensity for sufficiently low temperatures is explainable by the magnon BEC theory. However, the magnetization for $H>H_{\rm g}$ is the slightly convex function of $H$ \cite{Shiramura}, as observed in low-dimensional spin systems. This behavior disagrees with the magnon BEC theory.  \par
	We summarize the phase transition temperature $T_{\rm N}(H)$ and field $H_{\rm N}(T)$ in Fig.5, in which the data obtained from the previous magnetic measurements are also plotted. Because the transition points determined from the present neutron scattering experiments and the magnetic measurements lie almost on the same line, they are consistent with each other. This supports the prediction of the magnon BEC theory that the magnetization exhibits a cusplike minimum at the transition temperature \cite{Nikuni}. \par
\begin{figure}[tbp]
  \begin{center}
    \epsfigure{file=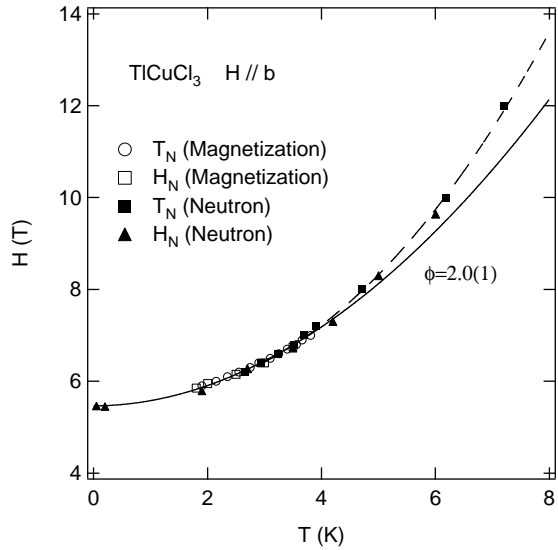,width=80mm}
  \end{center}
  \caption{The phase boundary in TlCuCl$_{3}$ for $H{\parallel}b$ determined from the results of the temperature variation (closed rectangles) and field variation (closed circles ) of the Bragg peak intensity. Open circles and rectangles denote the transition points determined from the previous magnetization measurements. The dashed line is a guide for the eyes. The solid line denotes the fit by eq. (1) with $(g/2)H_{\rm c}(0)=5.7(1)$ T and ${\phi}=2.0(1)$.}
  \label{fig:5}
\end{figure}
	The phase boundary can be expressed by the power law of eq. (1). In the previous magnetic measurements, the lowest temperature was 1.8 K. The low-temperature data contribute greatly to the determination of the exponent $\phi$. Therefore, adding the present data down to 0.05 K, we reevaluate the value of $\phi$. We fit eq. (1) to the data for $T<4$ K, which is lower than half the gap temperature $\Delta/k_{\rm B}=7.5$ K. The best fit is obtained with $H_{\rm g}=5.54$ T and $\phi =2.0(1)$, which is slightly smaller than the previous value of $\phi =2.2$. The exponent obtained, $\phi =2.0(1)$, is somewhat larger than the $\phi =1.5$ predicted by the magnon BEC theory. The difference between these values may be attributed to the fluctuation effect which is disregarded in the HF approximation. \par
	In conclusion, we have presented the results of neutron elastic scattering experiments performed on the spin gap system TlCuCl$_3$ in high magnetic fields along the $b$-axis. The field-induced transverse N\'{e}el ordering was clearly observed for $H>H_{\rm g}\approx 5.5$ T. The phase boundary can be described by the power law as predicted by the magnon BEC theory. However, the exponent obtained, $\phi=2.0$, is somewhat larger than the predicted value of $\phi=1.5$.  \par
The authors would like to thank T. Nikuni and M. Oshikawa for stimulating discussions. H. T., A. O. and T. K. express their gratitude to the Hahn-Meitner Institute for their kind hospitality during the experiment.  \par

\end{document}